# Mid-infrared-perturbed Molecular Vibrational Signatures in Plasmonic Nanocavities

Rohit Chikkaraddy*[1], Angelos Xomalis[1], Lukas A. Jakob[1], and Jeremy J. Baumberg*[1]

[1] NanoPhotonics Centre, Cavendish Laboratory, Department of Physics, JJ Thompson Avenue, University of Cambridge, Cambridge, CB3 0HE, United Kingdom

**Abstract**
**Recent developments in surface-enhanced Raman scattering (SERS) enable observation of single-bond vibrations in real-time at room temperature. By contrast, mid-infrared (MIR) vibrational spectroscopy is limited to inefficient slow detection. Here we develop a new method for MIR sensing using SERS. This method utilizes nanoparticle-on-foil (NPoF) nanocavities supporting both visible and MIR plasmonic hotspots in the same nano-gap formed by a monolayer of molecules. Molecular SERS signals from individual NPoF nanocavities are modulated in the presence of MIR photons. The strength of this modulation depends on the MIR wavelength, and is maximized at the 6-12μm absorption bands of $SiO_2$ or polystyrene placed under the foil. Using a single-photon lock-in detection scheme we time-resolve the rise and decay of the signal in a few 100ns. Our observations reveal that the phonon resonances of $SiO_2$ can trap intense MIR surface plasmons within the Reststrahlen band, tuning the visible-wavelength localized plasmons by reversibly perturbing the nanostructure crevices. This suggests new ways to couple nano-scale bond vibrations for optomechanics, with potential to push detection limits down to single-photon and single-molecule regimes.**

**Introduction**

Optical detection methods in the mid-infrared regime (MIR, 3-15μm) with single-photon sensitivity have wide implications in astrophysics and molecular nanoscience. Molecules and polar dielectric systems have characteristic bond vibrations and phonon modes across MIR wavelengths[1–6]. For ultrasmall sample volumes, optical detection (or pumping) of these modes gives low signals and is challenging due to the weak far-field coupling of these vibrations and low quantum efficiencies of MIR detectors[7,8]. Fourier transform infrared spectroscopy methods with photoconductive detectors (MCT) remain the workhorse for MIR detection, but they are slow, often require cryogenic cooling, and cannot approach the quantum limit. Upconverting low-energy MIR photons to high-energy visible photons would significantly benefit from single-photon-sensitive semiconductor (CCD, CMOS) technologies[9–11]. However, the poor conversion efficiencies and small spatial overlap of MIR and visible photons pose significant challenges[12].

Recent developments have circumvented the limitations associated with optical diffraction at long wavelengths by using near-field tip scanning (s-SNOM and PTIR)[13,14] and far-field mid-infrared photothermal microscopy (MIP)[15]. s-SNOM still relies on MCT-based detection schemes but can overcome diffraction limits from near-field scanning tips. Near-field (PTIR) and far-field (MIP) photothermal methods instead utilise efficient detection in the visible. The modulated MIR laser beam changes the reflection/transmission of a visible beam due to thermal expansion, pressure waves, refractive index changes or Grüneisen changes in the medium, which are efficiently detected through lock-in methods[15]. Even though the visible detectors used are fast and efficient, the signals obtained in PTIR and MIP are limited by thermal diffusivities on millisecond timescales.

These challenges can be addressed by up-conversion which utilizes cavity optomechanical approaches for efficient MIR detection. When MIR light impinges on an optical resonator it can excite mechanical resonances which are read out optically, allowing measurement at room temperature with low noise[16–18]. Here the detection limits are set by optomechanical coupling strengths ($g$), proportional to the quality factor of the mechanical mode. However, the diffraction-limited size of such cavities limits $g$ to less than 1MHz and thus functions worse than conventional MCT based detectors. Intriguingly, the mechanical motion can now be replaced by vibrating bonds in molecules (Fig.1a), opening clear avenues for molecular optomechanics and photochemistry[19,20]. This landscape of detection speed and resolution towards single-photon and single-molecule sensitivity shows how these diverse detection methods compare (Fig.1b).

Here we develop a MIR-perturbed surface-enhanced Raman scattering (SERS) method which uses single-molecule-sensitive metal nanocavities. The system is constructed using gold nanoparticles (AuNP) on a thin foil of planar Au with vibrating molecules assembled in the gap formed between them (Fig.1c). The strong visible-light confinement in the nanogap provides enhanced (>$10^9$) Raman scattering from the molecules in the gap, acting as a near-field probe. In this detection scheme, MIR light is absorbed in molecular bonds on the foil significantly altering the Stokes and anti-Stokes Raman signals at visible wavelengths, which can easily be detected (Fig.1a). The interaction of light and matter in these sub-nm mode volumes allows extreme sensitivity to (in principle) single MIR photon with resolution down to a single-molecule (Fig.1b).

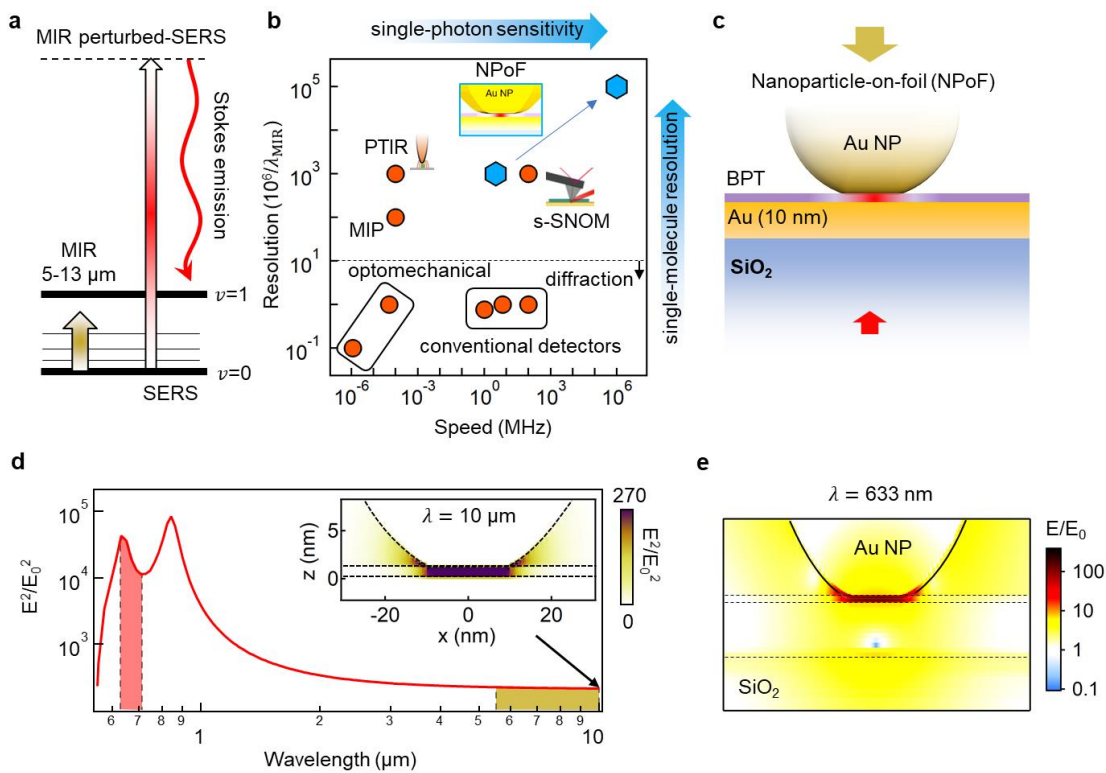

**Figure 1: Coupling MIR and visible light into plasmonic nanogaps.** (**a**) Energy diagram of MIR-perturbed SERS from molecule with ground ($v$=0) and excited ($v$=1) vibrational states. (**b**) Landscape of MIR detection speed and resolution comparing conventional and state-of the-art detection schemes with the NPoF devices. Faster detectors can distinguish single MIR photon arrival times in the detection stream. (**c**) Nanoparticle-on-foil (NPoF) constructed on thin metal film. (**d**) Simulated electromagnetic field enhancement *vs* wavelength in the center of NPoF gap. Inset shows enhanced light intensity at MIR $\lambda$=10µm. (**e**) Simulated electromagnetic field enhancement at 633nm used for SERS measurements.

## Results

Coherent electron oscillations coupled with light (plasmon polaritons) trap electromagnetic (EM) fields around metal nanostructures giving a resonant optical response in the visible and broad weaker optical response spanning from visible to MIR wavelengths[21–24]. While single metal nanoparticles do not provide sufficient field enhancement needed for robust single-molecule SERS, nanogap confinement improves this greatly. Here we exploit a multilayer nanoparticle-on-foil (NPoF) cavity that resonantly enhances the near-field at visible wavelengths in addition to giving a broad MIR optical response from lighting rod effects[25,26]. This structure consists of a faceted gold nanoparticle placed ~1.3nm above a thin Au film (10nm) deposited on a SiO$_2$ substrate (Fig.1c). The gap distance between the AuNP and Au-film is set by the monolayer height of biphenyl-4-thiol (BPT) molecules preassembled from solution onto the film before AuNP deposition. The resulting NPoF structure supports plasmonic $(lm)$ = (10) and (20) cavity resonances at 850nm and 650nm with E/E$_0$>100 (Fig.1e) [27–29]. This gives strong SERS and a broad uniform near-field enhancement across the MIR absorption wavelengths of 5-15µm (Fig.1d). The NPoF is designed for optimal spatial overlap of visible and MIR light which is vital for MIR-perturbed SERS.

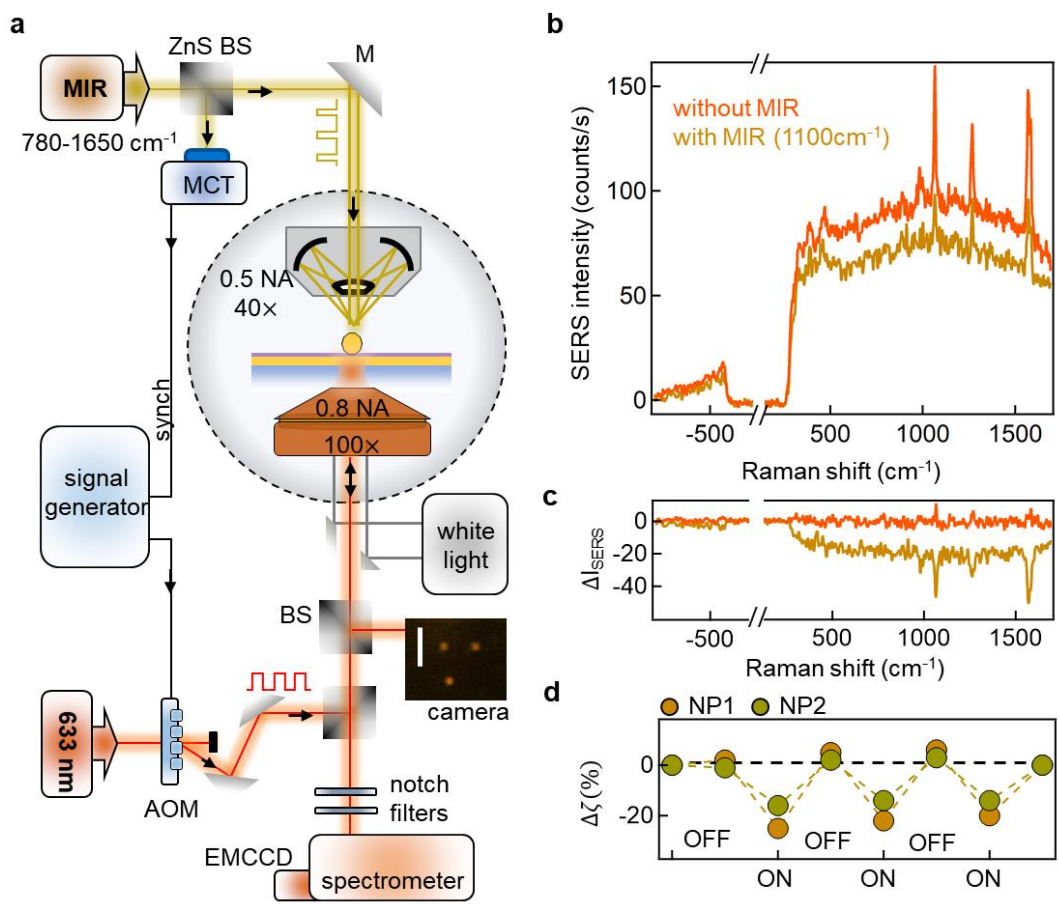

**Figure 2: MIR pump and visible SERS probe.** (**a**) Setup of MIR (800-1600 cm$^{-1}$) pump and visible (633 nm) probe beams illuminating an individual NPoF sample highlighted in black dotted circle. Acousto-optic modulator (AOM) for 633 nm SERS probe beam is synchronized with MIR pulses. SERS from NPoF sample is collected by a 0.8 NA 100× objective after filtering out probe laser with two notch filters. (**b**) SERS spectra from BPT self-assembled monolayer with and without the MIR pump at 1100 cm$^{-1}$. (**c**) Difference in the SERS amplitude induced by MIR, averaged over 3 scans on a single NPoF. (**d**) Perturbed change (Δζ %) in SERS intensity obtained by repeatedly switching the MIR beam ON/OFF for two different NPoFs.

To study the MIR-perturbed SERS from these cavities, we direct a tunable MIR pump beam (500μW average power) and 633nm SERS probe (150μW average power) onto individual NPoF cavities (Fig. 2a). The 633nm laser is focused through the $SiO_2$ substrate while the MIR pump is focused via a Cassegrain objective from the air-side, with estimated spot diameters on the sample of 1μm for 633nm and 20μm for the MIR beam at $\lambda$=10μm (Supporting Information, Fig.S1). Both beams are coaligned onto the sample and the back-scattered SERS from BPT molecules is collected though the $SiO_2$ substrate and routed to the spectrometer. The NPoF supports a unique dual configuration with metal-insulator-metal (MIM) gap mode at the AuNP-foil junction coupling to the insulator-metal-insulator (IMI) mode at the air-foil-glass interface, resulting in tightly confined MIMI modes which radiates SERS light predominately into the glass medium[30].

The NPoF cavities provide stable SERS signals upon laser illumination at 633nm, with characteristic BPT vibrational lines at 1080$cm^{-1}$ and 1585$cm^{-1}$ (Fig.2b).[31] The spectral intensity variation obtained from time-series spectra over a period of 30s from an individual NPoF cavity is <1% (Supporting Information Fig.S2). Upon irradiating with MIR light at 1100$cm^{-1}$, the SERS intensity is found to decrease by $\Delta\zeta$ >20% (Fig. 2b). This strong decrease in SERS signal is observed across all the vibrational lines of BPT as well as the Stokes background from the electronic Raman scattering (Fig. 2c). The observed intensity change immediately recovers once the MIR light is turned off (Fig. 2d).

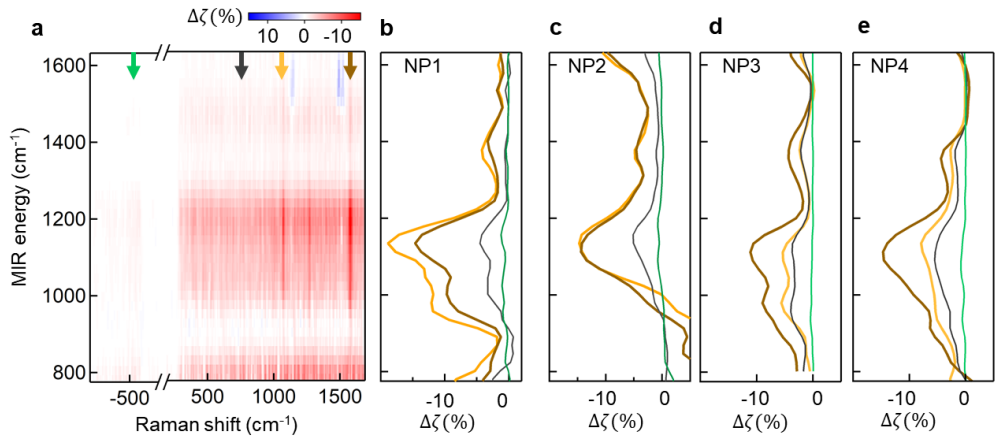

**Figure 3: MIR energy-dependent SERS change. (a)** Perturbed $\Delta\zeta$ (%) in the SERS signal from an individual NPoF when scanning the MIR frequency. $\Delta\zeta$ is normalized by the reference SERS before and after MIR illumination from the same NPoF cavity. **(b-e)** Perturbed $\Delta\zeta$ (%) extracted at the BPT vibrational lines at: 1580 $cm^{-1}$ (yellow), 1080$cm^{-1}$ (brown), Stokes ERS (grey) and anti-Stokes ERS (green), indicated by arrows in (a). Analysis from three other NPoFs is shown in **(c-e)**. (Unprocessed spectra in Supporting Information, Fig.S3, S4)

To understand the MIR energy-dependence we collect SERS spectra while tuning the MIR energy between 800-1600 cm$^{-1}$ (in steps of 20 cm$^{-1}$). SERS spectra are also collected both before and after the sample is illuminated with MIR light for reference. Scans with large variations (>30%) in the SERS spectra before and after MIR illumination either due to the alignment drift or diffusion of Au-adatoms in NPoF gaps are not considered [32–34]. Perturbed changes Δζ in SERS Stokes and anti-Stokes signals upon tuning the MIR illumination energy (Fig.3a) show a broadband response. Line profiles extracted from BPT at two different vibrational lines (1080cm$^{-1}$ and 1585cm$^{-1}$, Fig.3b-e) shows their maximum decrease occurs when the MIR is tuned around 1100cm$^{-1}$. Electronic Raman signals extracted from the Stokes background exhibit similar line profiles, with a lower magnitude (grey). This characteristic MIR perturbed peak at 1100cm$^{-1}$ corresponds to the SiO$_2$ phonon absorption (see below).

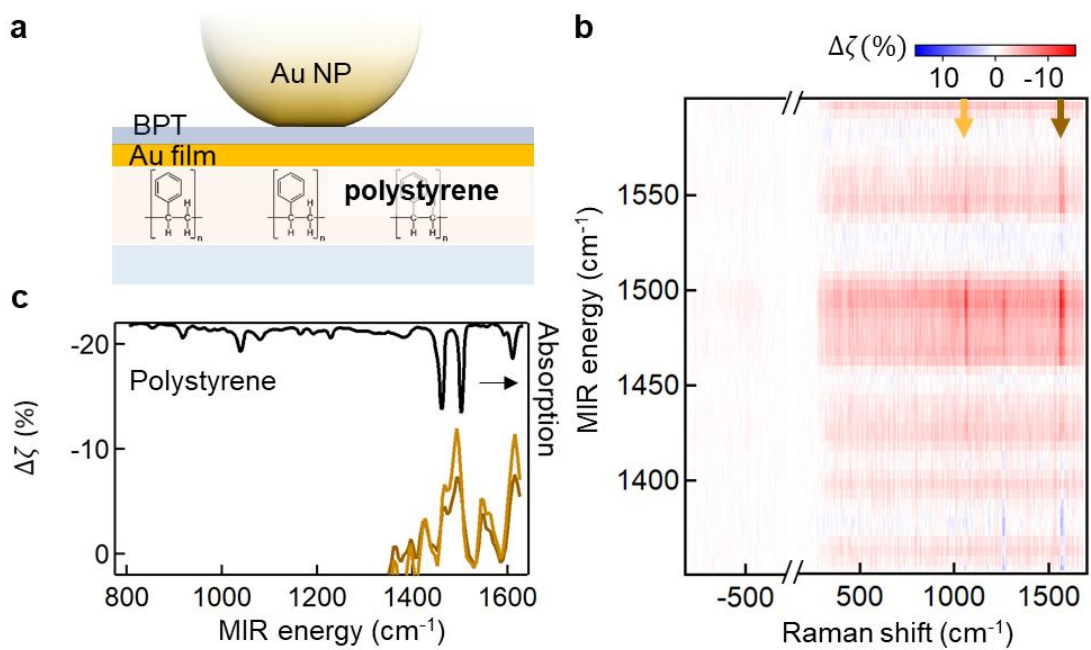

**Figure 4: MIR-perturbed SERS for polystyrene NPoF. (a)** NPoF sample with 500nm-thick polystyrene underneath the Au-foil. **(b)** MIR perturbed Δζ (%) in the SERS signal from an individual polystyrene-NPoF when scanning the MIR frequency. **(c)** Δζ (%) extracted at the BPT vibrational lines at: 1580 cm$^{-1}$ (yellow) and 1080cm$^{-1}$ (brown) from (b) and compared with the bulk polystyrene absorption (black).

To confirm the origin of MIR-perturbed signals from the influence of vibrations of the support material underneath the foil, we constructed NPoF systems with polystyrene replacing SiO$_2$ under the foil (Fig.4a). The MIR energy-dependence on polystyrene-NPoFs show a different spectral dependence and weaker signal intensity compared to SiO$_2$ samples. Here a smaller MIR frequency range is scanned with finer resolution of 5cm$^{-1}$. The MIR-perturbed SERS spectrum displays sharp peaks matching the vibrational absorption lines of bulk polystyrene, clearly indicating that the signal must originate from interactions with the material underneath the foil.

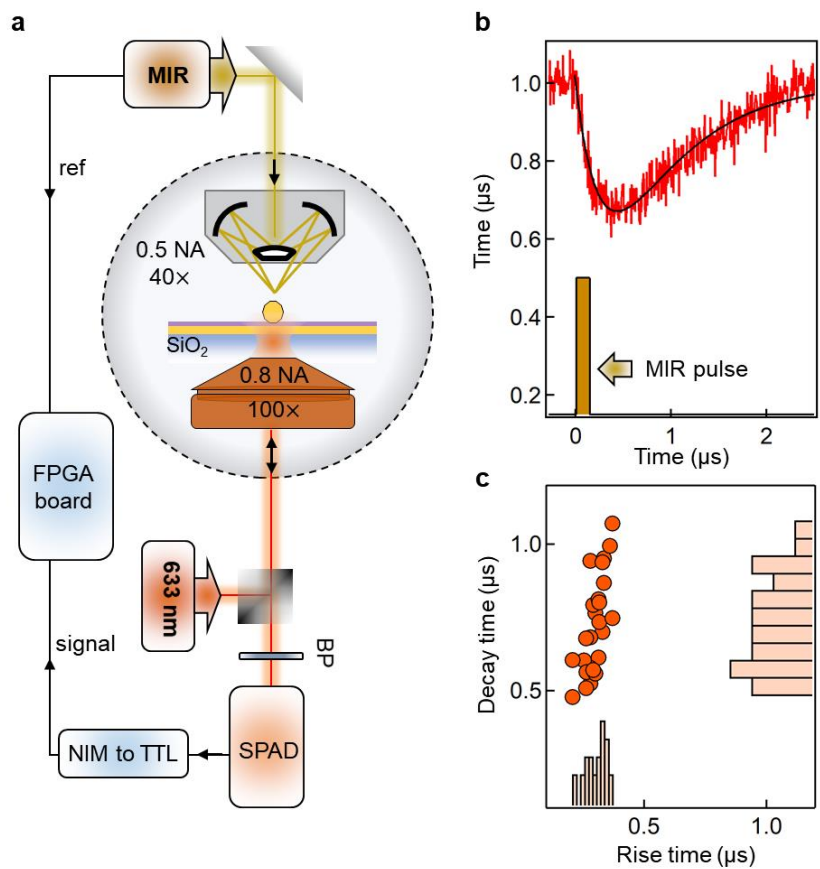

**Figure 5: Time-resolved single-photon lock-in. (a)** Modified microscope time-correlates trigger signals from MIR laser with SERS photons (filtered by band-pass BP) detected by single-photon SPAD in custom FPGA board. **(b)** Time-resolved MIR-perturbed SERS signal (red) from 100ns MIR pulse. Black curve is fit to extract rise and decay times. **(c)** Variation in rise time correlates with decay time, histograms of rise, decay times plotted on corresponding axes.

To characterize the dynamics of the MIR-perturbed SERS signal originating from the phonons underneath the foil, we develop a time-correlated single-photon lock-in method to time-resolve the signal. The Stokes part of the SERS signal is routed to a SPAD detector (Fig.5a). The arrival of each SERS photon is time correlated to the MIR trigger signal from the QCL. This provides a way to resolve the MIR-perturbed signal with time resolution of 100ps. The QCL is triggered at 0.32MHz with MIR pulses of width 100ns. The MIR-perturbed SERS rapidly decreases immediately after the MIR pulse (Fig.5b), with a timescale of ~300ns consistently obtained across multiple NPoF cavities (not limited by the MIR pulsewidth of 100ns). Subsequently the SERS signal recovers with a longer decay time of >500ns. This temporal response is fit with single exponential rise and decay times using experiments on >25 NPoF cavities. The rise time is narrowly distributed around 290±50ns whereas the decay time is more variable spanning 700±250ns (Fig. 5c). We find a positive correlation between the rise and decay times ($\tau_{\text{decay}} \sim 3.9 \tau_{\text{rise}}$), suggesting they are intrinsically linked to the origin of the MIR-perturbed SERS signature.

**Discussion**

The maximum decrease in SERS signal observed at MIR energies ~1100cm$^{-1}$ is consistent across different NPoFs, however the magnitude of signal varies between $\Delta\zeta$ =10-25%. The spectral response of the perturbed SERS does not match with Raman or IR vibrational lines of BPT. This indicates that MIR absorption in BPT is not the dominant contribution to the observed signal. Instead, this characteristic peak at 1100cm$^{-1}$ corresponds to the SiO$_2$ phonon absorption. This is also evidenced in the reflection dip that exhibits a typical Reststrahlen band[35,36] and confirmed by simulations of the MIR absorption at the Au-SiO$_2$ interface (Fig. 6a,b). Within a band between 900 and 1150 cm$^{-1}$, the real part of the SiO$_2$ dielectric function is negative (Re($\varepsilon$)<0). The reduced SERS signal must therefore arise from a decrease in near-field intensity of the 633nm probe, somehow caused by MIR excitation of this confined mode at the Au-SiO$_2$ interface. This results in a linear response with MIR power (Fig. 6c).

Direct heating of the Au interface from the 2mW average power MIR pump contributes only a minimal change of <1°C in temperature (Fig.6d,e), which is fully consistent with the unchanged anti-Stokes background of SERS signals observed. The change in refractive index of SiO$_2$ needed to account for a 20% decrease in SERS is rather high ($\Delta n$>0.2) for the pump powers used here (Supporting Information, Fig.S5), corresponding to temperatures >1000°C. As a result, simple thermal effects are not sufficient to account these observations. Further, conventional photothermal signals possess slow timescales (<ms)[37,38] as the induced deflection of visible light requires strong deformations of the substrate underneath. Thermal expansion of Au or SiO$_2$ for a 10K increase in local temperature is far too small to modulate the visible probe as required (Supporting Information, Table.S1). Similarly, reversible reconstruction of grain boundaries or polycrystallinity in the Au-foil on SiO$_2$ seems also unlikely to explain this MIR-perturbed SERS.

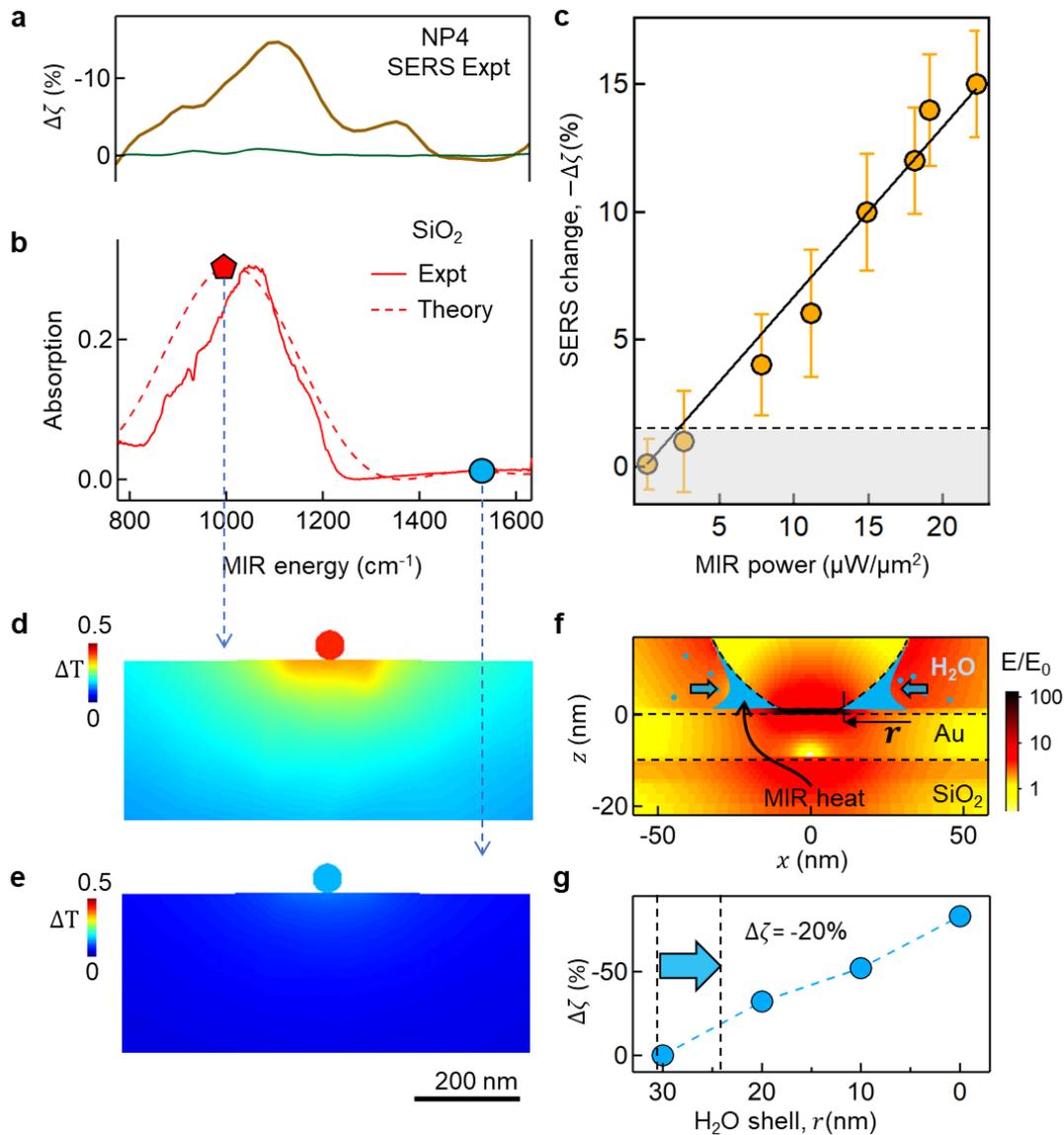

**Figure 6: SERS attenuation due to SiO₂ phonon absorption.** (**a**) Perturbed Δζ (%) in the SERS signal from an individual NPoF when scanning the MIR frequency. (**b**) Experimental and FDTD-simulated MIR absorption spectra of SiO$_2$-Au interface obtained in reflection geometry. (**c**) MIR (1100cm$^{-1}$) power dependent variation in the Δζ (%) of SERS signal. Solid line is linear fit, dashed gives detection noise baseline. (**d,e**) Steady state heat profile in NPoF simulated at (**d**) 1000cm$^{-1}$ and (**e**) 1500cm$^{-1}$. (**f**) Near-field enhancement of NPoF at 705nm (≡1580cm$^{-1}$ SERS line) surrounded by a 30nm-thick water shell (blue). (**g**) Simulated Δζ (%) for decreasing water shell thicknesses.

The decrease in SERS signal is thus attributed to a shift in the plasmon resonance wavelength perturbed by the modulation of refractive index directly around the AuNP (Supporting Information, Fig.S6). Exciting the SiO$_2$ Reststrahlen band shifts the (10) NPoF plasmon by ~1 nm and reduces the plasmon enhancement of SERS at $\lambda$=633nm. We can identify very few possible routes for this modulation, but possibilities can be either from nm-scale deformations in the NP surface or from the effects of a nanoscale shell of water in the crevices under the AuNP. This shell of water is always present for such nano-assemblies in ambient conditions, and extremely hard to remove due to the highly-acute crevice angle. Changing the crevice water shell width by <5nm is sufficient to induce a 20% decrease in SERS signal (Fig.6f,g). While the weak direct absorption of MIR light is insufficient to induce this, the situation is very different when the $SiO_2$ acts as a metal (from 900-1200cm$^{-1}$) which allows it to support surface-plasmon-polaritons (SPPs) that amplify the optical field near Au by >50. These MIR SPPs are excited only by scattering at the NP, leading to even higher fields directly in the crevices and thus heating trapped water in real time. Indeed, replacing the $SiO_2$ with $Si_3N_4$ (which has $Re(\varepsilon)>0$ throughout our spectral range) eliminates this effect, demonstrating the key role of resonant MIR SPPs. The absorbed energy from each MIR pulse is a hundred-fold larger than required to evaporate a 5nm-shell of water. This mechanism is also consistent with the sub-µs rise and decay times, which correspond to thermal diffusion times from the heated nanoparticle (Fig.6f).

Most dielectrics support vibrational mid-infrared-active phonon modes which interact with light and plasmons in the same fashion as described above[39,40]. These polariton modes are distributed across the MIR-visible regions and constrain the nanoscale geometries for producing upconverted SERS signals. Since MIR SPP excitation improves the MIR coupling into the gap, there is a trade-off between enhanced SERS upconversion and enhanced thermally- perturbed retuning of the plasmon resonances. Our work suggests that avoiding the substrate Reststrahlen band will be needed for observing SERS upconversion from molecules.

The efficiencies of MIR detection in this NPoF system are compared with state-of-the-art low-dimensional semiconductor heterostructures or graphene that have been implemented for THz detection in recent detection schemes[41–44]. From an application perspective, the relevant figure of merit is the noise equivalent power (NEP), which corresponds to the lowest detectable power in 0.5s integration time. This is measured here as the MIR power-dependent perturbation to the SERS signal (Fig.6c), however most of the incident MIR is reflected by the Au-foil. Given the 100nm$^2$ cross-section of NPoFs at MIR frequencies, the NEP is estimated to be 0.1nW Hz$^{-0.5}$, which is close to state-of-the-art detectors. Theoretically the noise level is limited by photon shot noise in the visible laser, although in current experiments the noise is limited by the stability of the SERS signal. Light-driven diffusion of adatoms[32,45] and fluctuations[33] of defects in the metal nanoparticle contribute to significant variation in SERS intensities. There exists an opportunity to significantly improve the noise reduction by developing more robust nanocavity systems. Further we suggest further improvements in MIR detection by deterministically creating adatom picocavities with light[34,46].

In summary, we show how molecular SERS signals are modified by irradiating with MIR light across a wide spectral bandwidth from 5.8 to 12µm (24 to 51 THz, 800 to 1700cm$^{-1}$). Our observations reveal that phonon resonances of the SiO$_2$ substrate trap intense MIR SPPs in the Reststrahlen band, which can temporarily retune the localised plasmons by perturbing the outer 5nm-thick shells of water in the nanostructure crevices. This results in strong reductions in SERS intensity, but could also be used in other ways, for instance for tuning plasmons in real time, as well as for exciting the NPoM in the MIR through SPP waveguides or antenna coupling. This suggests new ways to access nano-scale chemical imaging[3], MIR photothermal bolometers[47], photoacoustic microscopy[48] and optomechanics[49].

**Materials and Methods**

**Sample preparation.** To prepare the thin mirror, we deposit 10nm of Au on a clean SiO$_2$ cover slip (150µm thick) with a deposition rate of 0.5 Å/s (Moorfield nanoPVD-T15A thermal evaporator). The Au-coated SiO$_2$ substrates are dipped into a 1 mM solution of biphenyl-4-thiol (BPT, Sigma Aldrich, 97%) in anhydrous ethanol (Sigma Aldrich, <0.003% H2O) for 12h resulting in self-assembled molecular monolayers (SAMs). For NPoF optical cavities, 80nm faceted NPs (BBI Solutions) are deposited directly onto the BPT-assembled Au-coated SiO$_2$ substrates. The deposition time is kept below 30s, resulting in well-dispersed NPs. Lastly, the samples are rinsed thoroughly with double distilled water to remove the excess AuNPs.

**Experimental setup.** All SERS and MIR spectroscopy measurements are performed in a custom-built dual-channel microscope. For SERS, a spectrally-filtered 633nm diode laser (Matchbox, Integrated Optics) with 150µW/µm$^2$ power on the sample is used as a probe and is filtered with two notch filters before routing it to a Shamrock i303 spectrograph and a Newton EMCCD. The 633nm light is focused onto the sample with the aid of a x100 0.8 NA long working distance microscope objective. For imaging, the reflected light collected through the same objective lens is directed to a camera (Lumenera Infinity3-1). For the MIR light source, a LaserTune IR source (Block) with wavelength range of 5.4 - 13µm is used (1635 - 780 cm$^{-1}$) and maximum average output of 500 µW (~2x4 mm collimated) with 5% duty cycle. The pump (MIR light) is co-aligned with the probe (visible light) using a 0.4 NA Cassegrain objective lens. For MIR detection, an external mercury-cadmium-telluride (MCT) IR detector is used along with a ZnSe beam-splitter and is synced with the AOM to modulate the 633nm diode laser. This improves the pump and probe pulse temporal overlap by matching the repetition rate and pulse widths. The sample is placed on a fully automated motorized stage (Prior Scientific H101) which is controlled with code written in Python.

For single-photon time-correlated measurements, arrival times of all photons at the detector (Micro Photon Devices PDM $PD-100-CTD) and reference signals (MIR laser trigger) are

continuously recorded by a time-to-digital converter on a field-programmable gate array board. Comparing the photon timestamps with the reference signal allows recreating the periodic perturbation of the SERS signal by the MIR laser in time, integrated over millions of modulation cycles. This single-photon lock-in detection scheme is described in more detail elsewhere[49].

**Acknowledgements**

We acknowledge support from European Research Council (ERC) under Horizon 2020 research and innovation programme THOR (Grant Agreement No. 829067) and POSEIDON (Grant Agreement No. 861950). We acknowledge funding from the EPSRC (Cambridge NanoDTC EP/L015978/1, EP/L027151/1, EP/S022953/1, EP/P029426/1, and EP/R020965/1). R.C. acknowledges support from Trinity College, University of Cambridge.

**Author Information**

**Corresponding Author**

* Dr Rohit Chikkaraddy, rc621@cam.ac.uk
* Prof Jeremy J Baumberg, jjb12@cam.ac.uk

**Author Contributions**

R.C and J.J.B conceived and designed the experiments. R.C. performed the experiments with input from A.X and L.A.J. R.C. carried out the simulation and the analytical modelling with input from A.X. R.C. and J.J.B. analysed the data. R.C. and J.J.B. wrote the manuscript with input from all authors.

**Conflict of interest**

The authors declare no competing financial interest.

**Supporting Information**. A Supporting Information document is also provided, with additional images and information. Source data can be found at https://